\documentclass[conference,10pt]{IEEEtran}

\usepackage{graphicx}
\usepackage{amsmath}
\usepackage{amssymb}
\usepackage{amsfonts}
\usepackage{float}
\usepackage{color}
\usepackage{subfig}

\newtheorem{exemple}{Example}

\def\dotleq{\dot{\leq}}
\def\dotgeq{\dot{\geq}}

\title{Using Distributed Rotations for a Low-Complexity Dynamic Decode-and-Forward Relay Protocol}
\author{Charlotte Hucher and Parastoo Sadeghi\\
	The Australian National University, Canberra, ACT, 0200, Australia\\
        Email: \{charlotte.hucher,parastoo.sadeghi\}@anu.edu.au}

\begin{document}

\maketitle

\begin{abstract}
In this paper, we propose to implement the dynamic decode-and-forward (DDF) protocol with distributed rotations. In addition to being the first minimum-delay implementation of the DDF protocol proposed for any number of relays, this technique allows to exploit cooperative diversity without inducing the high decoding complexity of a space-time code.\\
The analysis of outage probabilities for different number of relays and rotations shows that the performance of this technique is close to optimal. Moreover, a lower-bound on the diversity-multiplexing gain tradeoff (DMT) is provided in the case of a single relay and two rotations. This lower-bound reaches the optimal DDF's DMT when the frame-length grows to infinity, which shows that even a small number of rotations is enough to obtain good performance.
\end{abstract}

\begin{keywords}
 dynamic decode-and-forward (DDF), relaying protocol, cooperative diversity, distributed rotations, outage probability, diversity-multiplexing gain tradeoff
\end{keywords}

\section{Introduction}

The last decade has witnessed a growing interest in wireless cooperative communications \cite{sendonaris03}-\nocite{nosratinia04}\cite{kramer05}. Indeed wireless nodes cannot always be equipped with several antennas due to size, cost or hardware limitations. But some diversity can still be exploited by considering the virtual multiple antenna array formed by several nodes of the network and using distributed transmission techniques.

Cooperative protocols have been classified into different families according to the processing performed at relays. One of these families is formed by the decode-and-forward (DF) protocols. In theory, these protocols could bring significant improvements in performance thanks to the noise deletion at relays. However, they are limited by the source-relay capacities since relays have to decode the signals correctly to be able to forward them.

The dynamic decode-and-forward (DDF), proposed in \cite{azarian05}, is based on the idea that each relay should listen till it receives enough information to decode, and retransmit only then. It has been proven to outperform any amplify-and-forward (AF) protocol in terms of diversity-multiplexing gain tradeoff (DMT). However, this protocol is quite complex and providing a simple implementation is still an open problem.

Recently proposed in \cite{yang10}, distributed rotations is a new technique to exploit spatial diversity without adding the decoding complexity of a space-time code. In \cite{yang10}, the authors show that this technique is optimal in terms of DMT for the two-hop multiple-relay channel using an amplify-and-forward strategy.

In this paper, we propose to implement a low-complexity DDF protocol with distributed rotations and analyze the performance of this scheme in terms of outage probability. We also take a special interest in the case of a single relay and analyze its performance for a small number of rotations in terms of DMT. Finally we discuss the implementation of this protocol and study the impact of transmitting data blocks instead of single symbols.

\section{Channel model and background}
\subsection{Channel model}

We consider a cooperative network with one source $S$, a set of $N$ relays $R_1$ to $R_N$, and one destination $D$. All nodes are half-duplex and equipped with a single antenna.

\begin{figure}[t]
 \centering
 \includegraphics[width=.65\linewidth,clip]{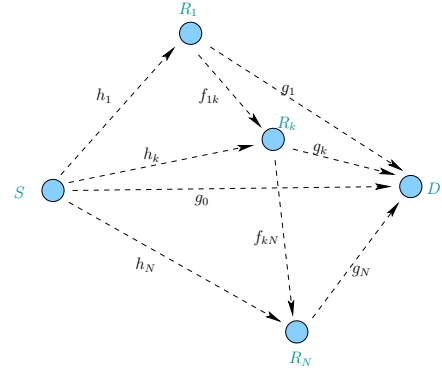}
 \caption{Relay channel model.}
 \label{channel_model}
\end{figure}

In this paper, notations of Fig. \ref{channel_model} are used. The $S\rightarrow D$, $S\rightarrow R_i$, $R_i\rightarrow R_k$ and $R_k\rightarrow D$ fading channel links are denoted by $g_0$, $h_i$, $f_{ik}$ and $g_k$, respectively, and follow complex Gaussian  distributions. The channel is assumed to be slow-fading. Thus $g_0$, $h_i$, $f_{ik}$ and $g_k$ can be considered as constant during the transmission of at least one frame of length $T$.

All receivers are assumed to have perfect channel state information (CSI), i.e. the destination $D$ knows the values of $g_0$ and all $g_k$, and each relay $R_k$ knows the values of $h_k$ and all $f_{ik}$.

The power is uniformly distributed between the source and the relays as our aim in this paper is not to optimize power allocation.

\subsection{Dynamic decode-and-forward}

The DDF was first proposed in \cite{azarian05} by Azarian \textit{et al}. 

The source transmits during the whole transmission frame. Each relay listens to signals transmitted by the source and possibly other relays till it is able to decode the message. It then retransmits this decoded version. For each relay, the length of the first phase is thus variable and changes dynamically depending on the channel realization and the noise.

The authors show in \cite{azarian05} that in the single relay case, if the outage criteria is considered, the relay is able to decode after $T_1$ timeslots:
\begin{equation}
T_1 = \min \left\lbrace T , \left\lceil \frac{TR}{\log(1+\rho|h_1|^2)} \right\rceil \right\rbrace,
\end{equation}
where $\rho$ is the signal-to-noise ratio (SNR) and $R$ is the spectral efficiency in bits per channel use (bpcu).

Let $r$ and $d(r)$ be the multiplexing and diversity gains, respectively. $r$ and $d(r)$ are defined by $R = r\log(\rho)$ and
\begin{displaymath}
d(r) = \lim_{\rho\rightarrow\infty} \frac{\log(p_{out}(r\log(\rho))}{\log(\rho)}.
\end{displaymath}

In \cite{azarian05}, the authors calculated the DMT of the DDF and proved that it is better than for any known AF or DF protocol, and even optimal for $r\leq\frac{1}{N+1}$:
\begin{equation}
d(r) = \left\lbrace \begin{array}{ll} 
         (N+1)(1-r)            & \text{ if }  0 \leq r \leq \frac{1}{N+1}, \\
	 1+\frac{N(1-2r)}{1-r} & \text{ if }  \frac{1}{N+1} \leq r \leq \frac{1}{2}, \\
	 \frac{1-r}{r}         & \text{ if }  \frac{1}{2} \leq r \leq 1.
                      \end{array}\right.
\end{equation}
This DMT is obtained not only by studying the asymptotic behavior of the outage probability when the SNR grows to infinity, but also considering an infinite frame. Thus in practice, this DMT is not achievable, but can be approached as close as desired.

This protocol is quite hard to implement in practice: indeed relays (except for the first one to decode) have to take into account the signals broadcasted by both source and other relays in order to decode as soon as possible. Moreover, at each time slot, a different number of users are transmitting. The construction of a space-time code is thus a complex problem, since it should be adapted to different numbers of antennas.

For the one-relay case, several implementation techniques have been introduced in \cite{kumar09,plainchault10}. In the case of several relays, the only proposed implementation to the authors' knowledge is based on space-time coding \cite{elia09}.

\subsection{Distributed rotations}

Distributed rotations is a new technique to exploit spatial diversity that was recently introduced in \cite{yang10}.

Let $L$ be the number of considered rotations. We define the set of used angles as
\begin{equation}
 \mathcal A_L = \left\lbrace 0, \frac{2\pi}{L}, \dots, \frac{2(L-1)\pi}{L} \right\rbrace,
\end{equation}
and the set of rotations as
\begin{equation}
 \varTheta_L = \left\lbrace e^{i\theta}, \theta\in\mathcal A_L \right\rbrace.
\end{equation}

A distributed rotation array of dimension $N$ is an array of $N$ rotations taken from $\varTheta_L$. We define the matrix $\mathbf R = (r_{kt})$ of size $N\times T$, $T\geq L^N$, such that all possible distributed rotation arrays appear as a column of $\mathbf R$. If $T=L^N$, then $\mathbf R$ is the set of all possible combinations of the $L$ rotations by $N$ relays.

In \cite{yang10} the authors apply these distributed rotations to a two-hop multiple-relay channel and claim that it is optimal in terms of DMT.

\begin{figure}[h!t]
 \centering
 \subfloat[$g_1+e^{i\theta}g_2$]{\includegraphics[clip, scale=0.6]{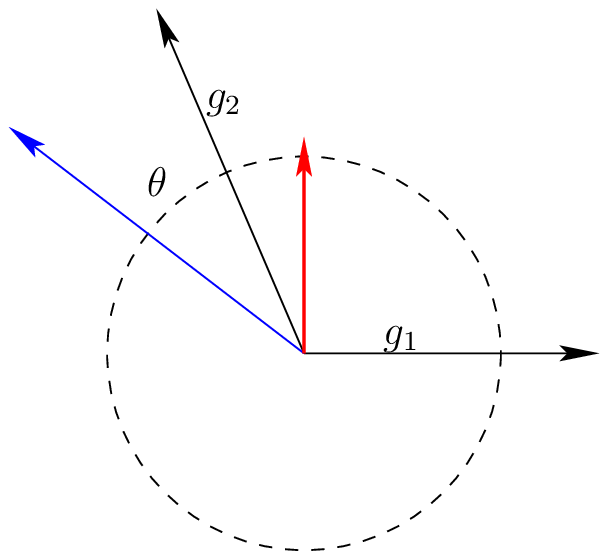}}
 \hfil
 \subfloat[$g_1+e^{i(\theta+\pi)}g_2$]{\includegraphics[clip, scale=0.6]{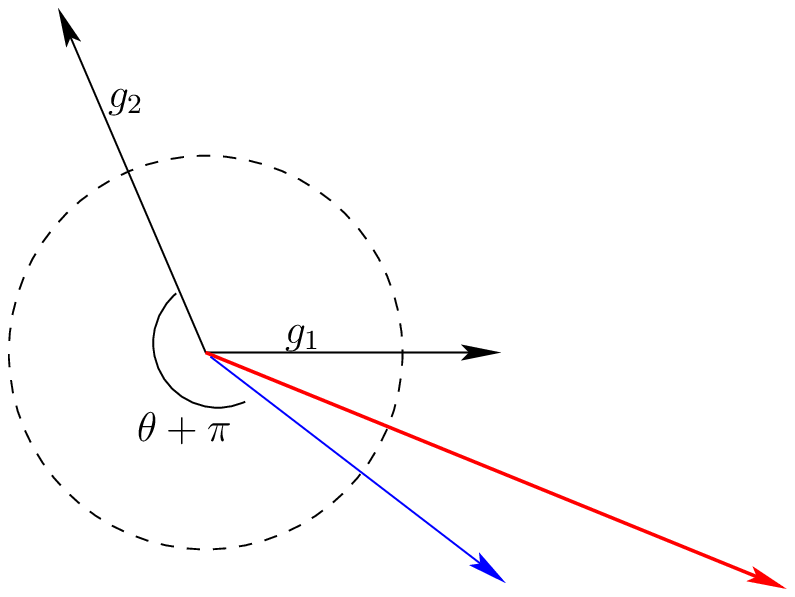}}
 \caption{Destructive and constructive combinations of the channels.}\label{fig:combi}
\end{figure}

In particular, this technique allows to exploit space diversity. Indeed, distributed rotations applied to the different channel links can be seen as random channel alignments. Let us consider the following example:

\begin{exemple}
Let $g_1$ and $g_2$ be two complex Gaussian distributed channel links. Let us consider an even number of rotations $L$: if $\theta\in\mathcal A_L$, then $(\theta+\pi)\in\mathcal A_L$ too. If the source transmits only through channel $g_1$ and if $g_1$ is subject to a strong fading, then information will be lost. But if combinations of the two channels with different rotations are considered, at least one of the combinations will be good in terms of channel alignment. Indeed if the combination $g_1+e^{i\theta}g_2$ is destructive, then $g_1+e^{i(\theta+\pi)}g_2$ is constructive (see Fig. \ref{fig:combi}). Thus the information sent through this channel will be lost iff both channels $g_1$ and $g_2$ are subject to a strong fading. Thus a diversity order of $2$ can be achieved in this case.
\end{exemple}

\section{Distributed Rotations for the Dynamic Decode-and-Forward Protocol}

The idea in this paper is to apply distributed rotations to the dynamic decode-and-forward protocol in order to provide an efficient transmission scheme with a relatively low decoding complexity.

\subsection{System model}

Let $\mathbf s$ be a sequence of $T$ information symbols. Let $\mathbf x$ be the sequence of $T$ coded symbols at the source: $\mathbf x = \mathbf C \mathbf s$, where the code matrix $\mathbf C$ is full-ranked with non-zero elements.

The source is always transmitting.
At each time slot, only the relays which have been able to decode can forward the information. They then transmit the same symbol as the source, but rotated by a predefined rotation. If at time $t$, relay $R_k$ has been able to decode, then it transmits $r_{kt} x_t$.

Let us consider with no loss of generality that relays are ordered, so that relay $R_1$ is the first one to decode and $R_N$ the last one. Let $T_k$ be the length of the listening phase for relay $R_k$:
\begin{displaymath}
 T_1 \leq T_2 \leq \dots \leq T_N.
\end{displaymath}
Then, for $T_j < t \leq T_{j+1}$, received signals at the destination and relay $R_i$ are
\begin{equation}
 y_t^d = \sqrt{\rho} \frac{1}{\sqrt{1+j}} \left(g_0 + \sum_{k=1}^j r_{kt} g_k\right) x_t + w_t,
\end{equation}
and 
\begin{equation}
 y_t^i = \sqrt{\rho} \frac{1}{\sqrt{1+j}} \left(h_i + \sum_{k=1}^j r_{kt} f_{ki}\right)  x_t + w_t^i,
\end{equation}
respectively, where $w_t$ and $w_t^i$ are additive white Gaussian noise (AWGN) with unit variance. The factor $\frac{1}{\sqrt{1+j}}$ is added so that the total power of transmission is constant.

From a receiver point of view (destination or relay), this is equivalent to a fast fading single-input single-output (SISO) channel, which allows a simple decoding. For $T_j < t \leq T_{j+1}$, let $G_{t} = g_0 + \sum_{k=1}^j r_{kt} g_k$ and $H_{it} = h_i + \sum_{k=1}^j r_{kt} f_{ki}$ be the equivalent fast fading channels. The rotations are fixed and used in a fixed order for each relay, which is known by all the receivers. Thus relay $R_i$ and destination $D$ can compute $H_{it}$ and $G_t$, respectively, assuming that they know whether each relay is transmitting or not. This information can be broadcasted by each relay using a limited amount of bits.

\subsection{Outage probability}

In the outage probability analysis, each relay starts transmitting as soon as it has been able to decode, according to the outage criteria. Indeed, according to Shannon's theorem, if the channel seen by the relay is in outage, then no detection is possible without error. On the contrary, if the channel is not in outage, then there exists an error-correcting code that allows to obtain an error probability as small as desired. Let $p(T_i,R)$ be the probability of the size of the listening phase of relay $R_i$ being $T_i$ for data rate $R$.

The probability $p(T_1,R)$ of the first relay is independent of the other relays, but it is not the case anymore for the next relays. Indeed, subsequent relays have to listen to both source and previous relays, and decode from this combination of signals. The probability $p(T_i,R|T_1,\dots,T_{i-1})$ thus depends on the decoding times of previous relays:
\begin{align}
 & p(T_i,R|T_1,\dots,T_{i-1}) \\
 & \nonumber = \Pr\left\lbrace \sum_{j=0}^{i-1} \sum_{t=T_j+1}^{T_{j+1}} \log \left( 1 + \frac{\rho}{1+j} |H_{it}|^2 \right) \geq TR, \right.\\
 & \nonumber \left. \hspace{1cm} \sum_{j=0}^{i-1} \sum_{t=T_j+1}^{T_{j+1}-1} \log \left( 1 + \frac{\rho}{1+j} |H_{it}|^2 \right) < TR \right\rbrace
\end{align}
where we let $T_0 = 0$. For $T_j+1<t\leq T_{j+1}$, the time-varying channel $H_{it}$ is the sum of $j+1$ terms corresponding to the source and the $j$ previous relays.

The outage probability of the DDF protocol implemented with distributed rotations is then
\begin{align}
& p_{out}(R)  = \sum_{T_1\leq\dots\leq T_N} \left( \prod_{i=1}^N p(T_i,R|T_1,\dots,T_{i-1}) \right)\\
\nonumber  & \times \Pr \left\lbrace \left.\sum_{j=0}^{N} \sum_{t=T_j+1}^{T_{j+1}} \log \left( 1 + \frac{\rho}{1+j} |G_t|^2 \right) < TR \right| T_1,\dots,T_N \right\rbrace
\end{align}
where we let $T_{N+1} = T$.

\subsection{The case of a single relay and 2 rotations}

\subsubsection{Outage probability}

In the one-relay case, the relay attempts to decode the information from signals broadcasted by the source only. Thus, we can express the probability of the listening phase of the relay as
\begin{align}
p(T_1,R) =  & \Pr\left\lbrace  T_1\log \left( 1+\rho|h_1|^2 \right) \geq TR, \right.\\
\nonumber & \hspace{1cm} \left. (T_1-1) \log \left( 1+\rho|h_1|^2 \right) < TR \right\rbrace.
\end{align}

If only $L=2$ rotations are considered, then the outage probability of our DDF scheme is
\begin{align}
p_{out}(R)  = & \sum_{T_1=1}^T p(T_1,R) \Pr \Bigg\lbrace T_1 \log \left( 1 + \rho |g_0|^2 \right) \\
\nonumber  & \hspace{1cm} + \sum_{t=T_1+1}^{T} \log \left( 1 + \frac{1}{2} \rho |g_0 + r_t g_1|^2 \right) < TR  \Bigg\rbrace
\end{align}
where $\forall t \in \{1,\dots,T\}$, $r_t\in \varTheta_2 = \{1,-1\}$.

For a given channel realization, the transmitting phase of the relay lasts $T-T_1$ time slots. Since there is $L=2$ rotations, during half of the time the relay uses one of these rotations, and during the other half it uses the other one. If $T-T_1$ is an odd number, we can assume without loss of generality that the second rotation is used one more time. Then, we can write
\begin{align}
\nonumber p_{out}(R)  = & \sum_{T_1=1}^T p(T_1,R) \Pr \Bigg\lbrace T_1 \log \left( 1 + \rho |g_0|^2 \right) \\
                     & \hspace{.5cm} + \left\lfloor\frac{T-T_1}{2}\right\rfloor \log \left( 1 + \frac{1}{2} \rho |g_0 + g_1|^2 \right) \\
\nonumber            & \hspace{.5cm} + \left\lceil\frac{T-T_1}{2}\right\rceil \log \left( 1 + \frac{1}{2} \rho |g_0 - g_1|^2 \right) < TR  \Bigg\rbrace,
\end{align}
which, with $A = \left\lfloor\frac{T-T_1}{2}\right\rfloor$, can be upper-bounded by
\begin{align}\label{pout}
& p_{out}(R) \leq \sum_{T_1=1}^T p(T_1,R) \Pr \left\lbrace T_1 \log \left( 1 + \rho |g_0|^2 \right) \right. \\
\nonumber & \left. + A \log \left( 1 + \rho (|g_0|^2 + |g_1|^2) + \frac{1}{4} \rho^2 (|g_0|^2 - |g_1|^2)^2 \right) < TR  \right\rbrace.
\end{align}

\subsubsection{Diversity-multiplexing gain tradeoff (DMT)}

In order to compute a lower-bound on the DMT of the DDF protocol implemented with distributed rotations, we study the behavior of the outage probability when $\rho$ grows to infinity.

Let $u$, $v_0$ and $v_1$ be the exponential orders of $|h|^{-2}$, $|g_0|^{-2}$ and $|g_1|^{-2}$, respectively. Let $\doteq$, $\dotleq$ and $\dotgeq$ denote the behavior of the left term of the equality/inequality when the SNR grows to infinity. Using these notations, the asymptotic behavior of $p(T_1,R)$ can be expressed as
\begin{align*}
p(T_1,r\log(\rho)) & \doteq \Pr\left\lbrace  (T_1-1)\max(0,1-u) < Tr \right\rbrace \\
                   & \doteq \Pr\left\lbrace  1-\frac{T}{T_1-1}r < \min(1,u) \right\rbrace \\
                   & \doteq \rho^{-d^1(T_1,r)}
\end{align*}
where $(a)^+ = \max(0,a)$ and
\begin{equation}
d^1(T_1,r) = \left(1-\frac{T}{T_1-1}r\right)^+.
\end{equation}

The probability that $|g_0|^2=|g_1|^2$ is equal to $0$, thus the last term of the mutual information in (\ref{pout}) is dominant when $\rho$ grows to infinity and we can write
\begin{align}
& p_{out}(r\log(\rho)) \dotleq \sum_{T_1=1}^T \rho^{-d^1(T_1,r)} \Pr \Bigg\lbrace T_1 \max(0,1-v_0) \\
\nonumber & \hspace{2cm} + A \max(0,2-2v_0,2-2v_1) < Tr \Bigg\rbrace.
\end{align}

Let $\mathcal B$ be the two-dimensional region defined by the inequalities $T_1 \max(0,1-v_0) + A \max(0,2-2v_0,2-2v_1) \leq Tr$ and $0 \leq v_0,v_1 \leq 1$. Then,
\begin{equation}
d(T_1,r) \geq \inf_\mathcal B \{v_0+v_1\}.
\end{equation}

\begin{figure}
 \centering
 \subfloat[]{\includegraphics[clip, scale=0.25]{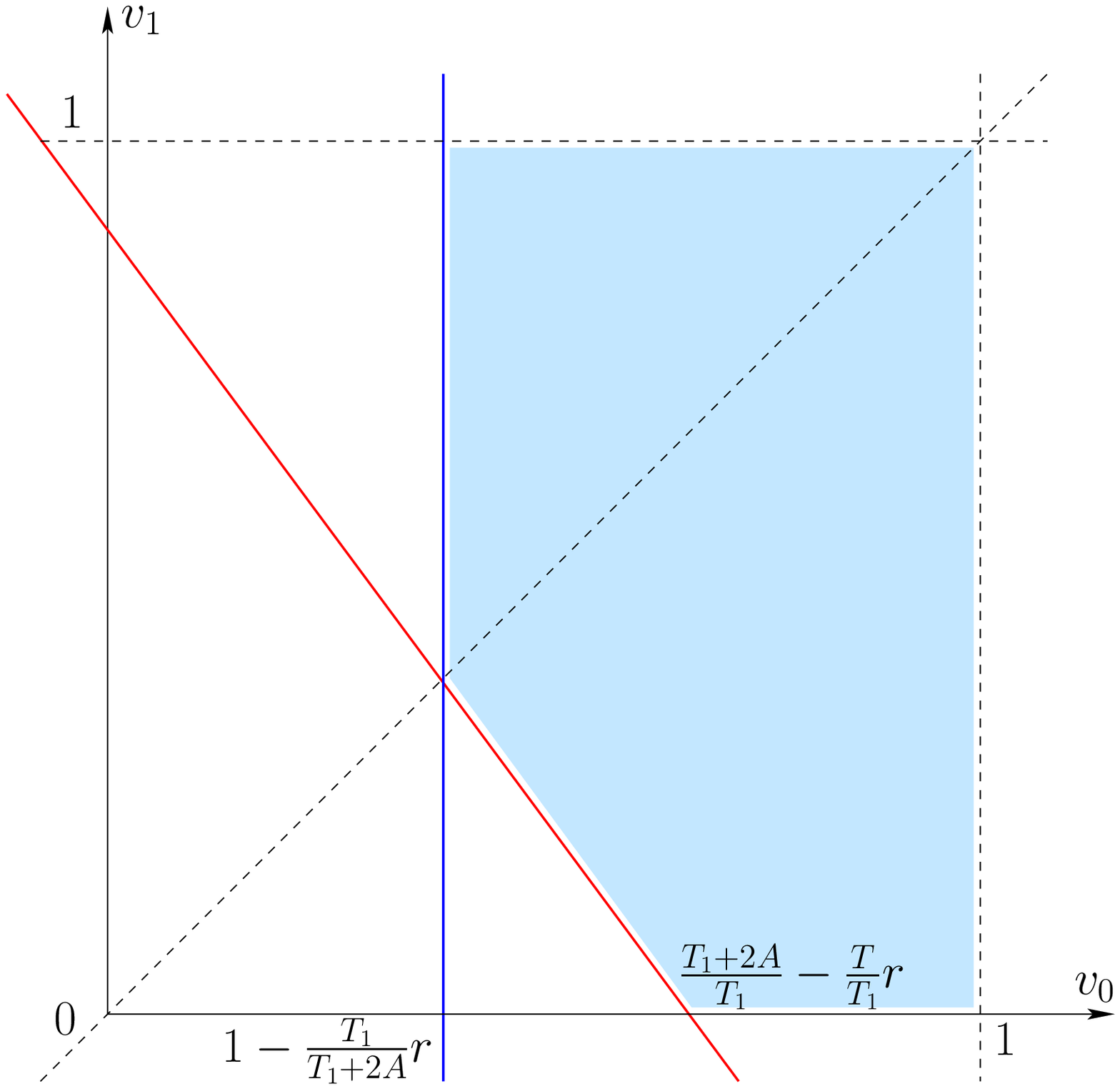}\label{dmt_region1}}
 \hfil
 \subfloat[]{\includegraphics[clip, scale=0.25]{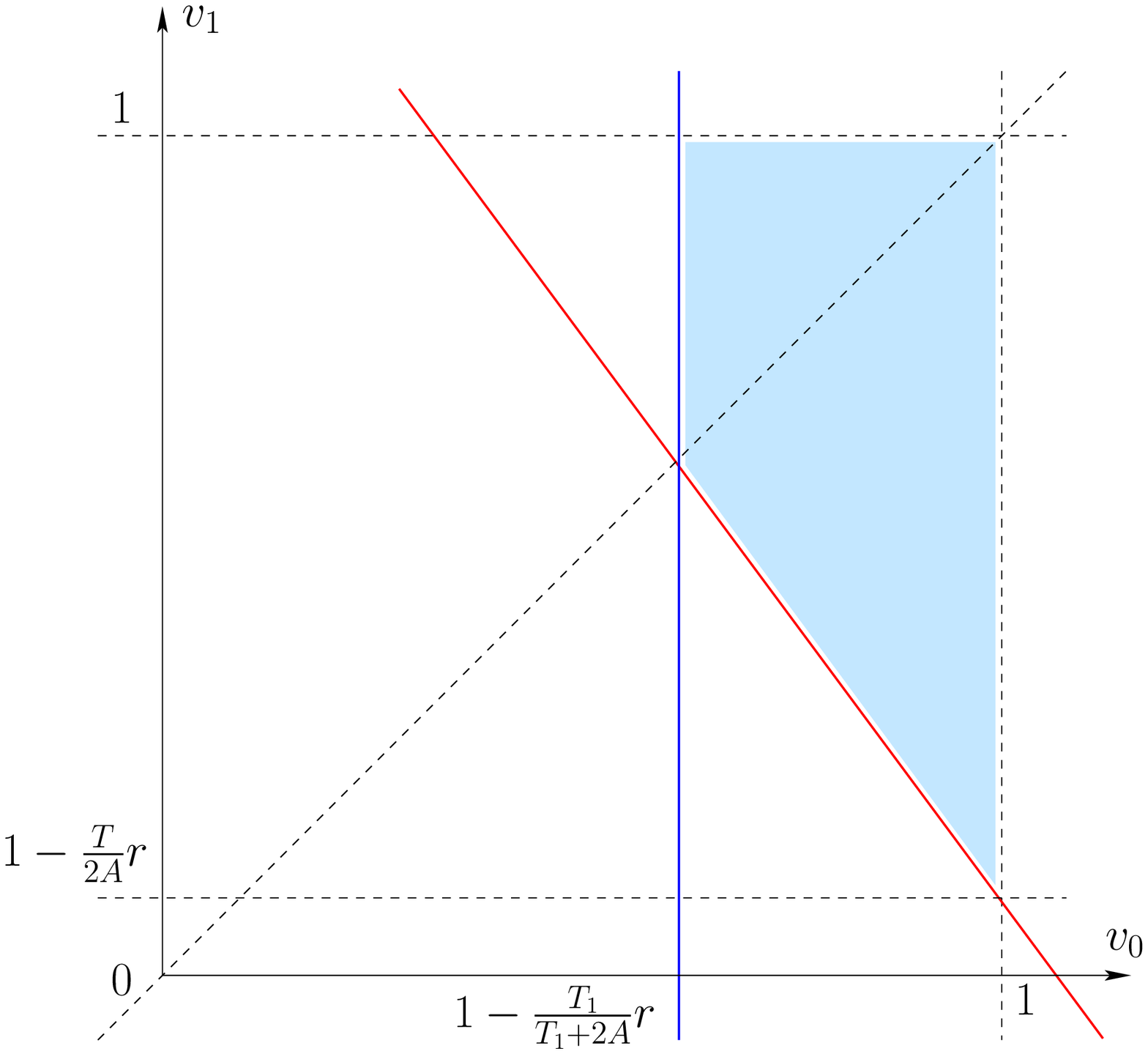}\label{dmt_region2}}
 \caption{Region $\mathcal B$.}\label{dmt_region}
 \centering
 \includegraphics[clip,width=.9\linewidth]{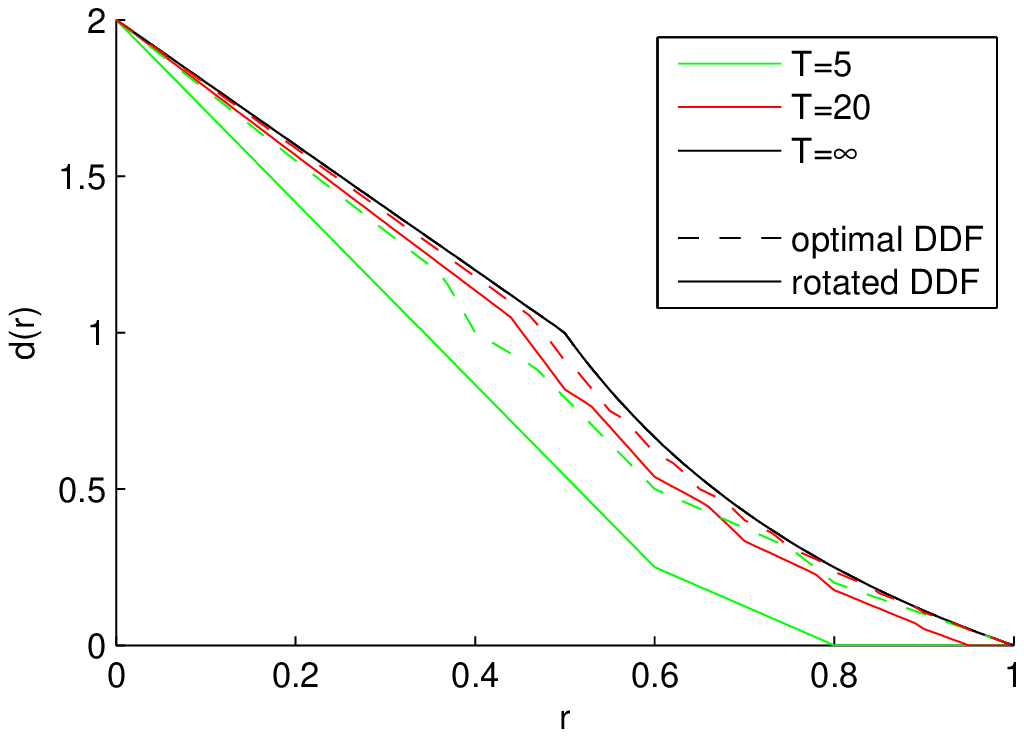}
 \caption{DMT of the DDF protocol for different frame-lengths.}\label{dmt}
\end{figure}

\begin{itemize}
 \item if $v_0 \geq v_1$, then $T_1 (1-v_0) + 2A (1-v_0) < Tr$ \\ $\Leftrightarrow 1-\frac{T}{T_1+2A}r < v_0$ (blue line on Fig. \ref{dmt_region})
 \item if $v_0 \leq v_1$, then $T_1 (1-v_0) + 2A (1-v_1) < Tr$ \\  $\Leftrightarrow T_1 v_0 + 2A v_1 > T_1 + 2A - Tr$ (red line on Fig. \ref{dmt_region})
\end{itemize}
Fig. \ref{dmt_region} representing the region $\mathcal B$ (in light blue) shows that
\begin{itemize}
 \item if $r \geq \frac{2A}{T}$ (see Fig. \ref{dmt_region1}), then
\end{itemize}
  \begin{equation}
   d(T_1,r) \geq \left\lbrace \begin{array}{l}
                            2\left(1-\frac{T}{T_1+2A}r\right)^+ \text{ if } T_1 \leq 2A \\
			    \frac{T_1+2A}{T_1}\left(1-\frac{T}{T_1+2A}r\right)^+ \text{ if } T_1 \geq 2A
                           \end{array}\right.,
  \end{equation}
\begin{itemize}
 \item and if $r \leq \frac{2A}{T}$ (see Fig. \ref{dmt_region2}), then
  \begin{equation}
   d(T_1,r) \geq \left\lbrace \begin{array}{l}
                            2\left(1-\frac{T}{T_1+2A}r\right)^+ \text{ if } T_1 \leq 2A \\
			    \left(2-\frac{T}{2A}r\right)^+ \text{ if } T_1 \geq 2A
                           \end{array}\right..
  \end{equation}
\end{itemize}

Finally in the single relay case, for two rotations, we can express the DMT of the DDF protocol as
\begin{equation}
 d(r) \geq \min_{T_1} d^1(T_1,r) + d(T_1,r).
\end{equation}

The lower-bound on the DMT is plotted in Fig. \ref{dmt}. One can see that even for a small number of rotations, the distributed rotations technique succeeds in asymptotically reaching the optimal DMT of the DDF protocol when the frame-length grows to infinity.

\section{Analysis of Performance}

In the following, outage probabilities of the DDF protocol implemented with distributed rotations have been obtained through Monte-Carlo simulations (averaging on both the channel realization and the rotation order) for different number of relays and rotations and a frame-length of $64$ symbols.

\begin{figure}[h!t]
 \centering
 \includegraphics[clip,width=\linewidth]{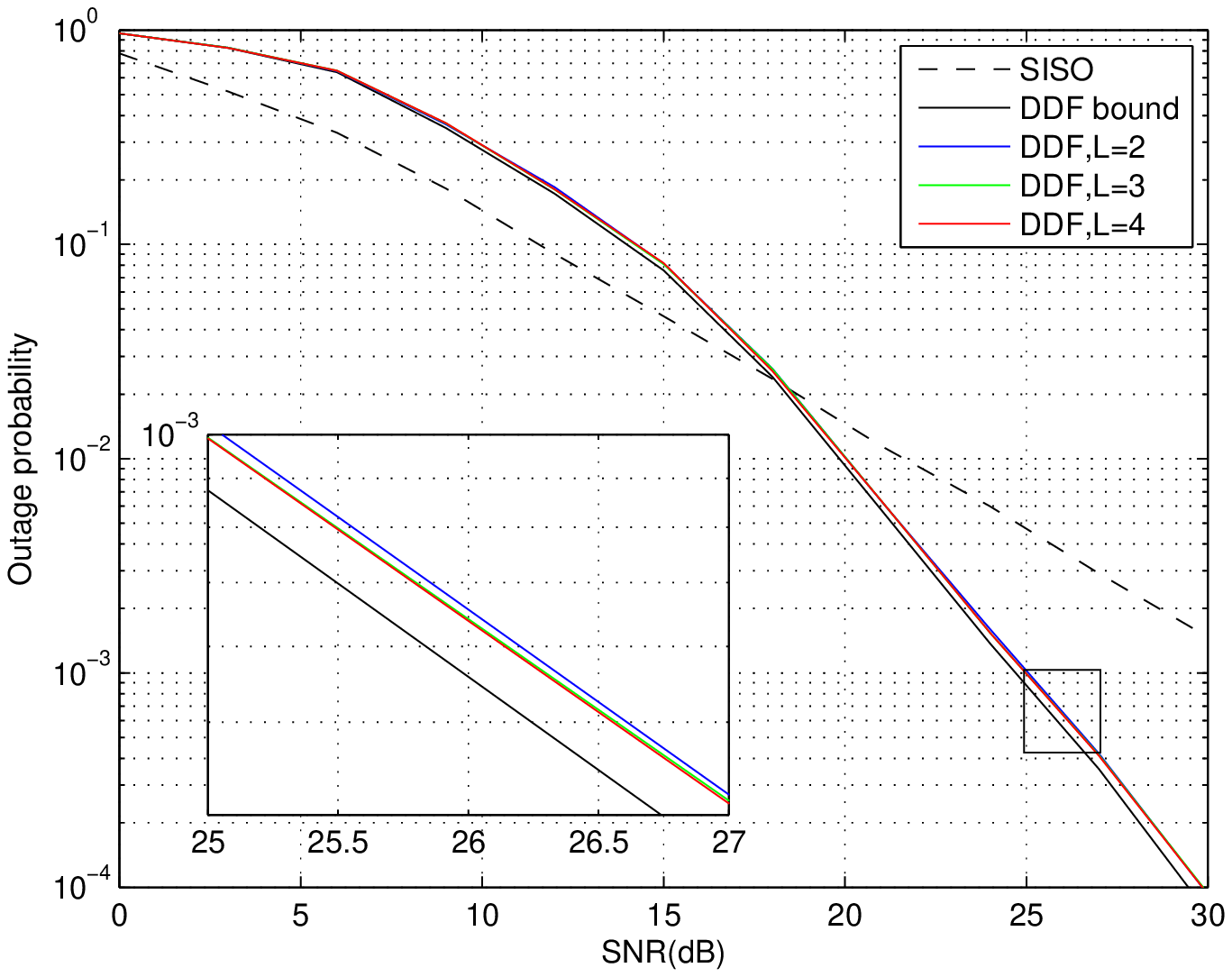}
 \caption{Outage probabilities of the DDF protocol implemented with different number of rotations for 1 relay with a frame-length of 64 symbols.}\label{pout_N1}
 \includegraphics[clip,width=\linewidth]{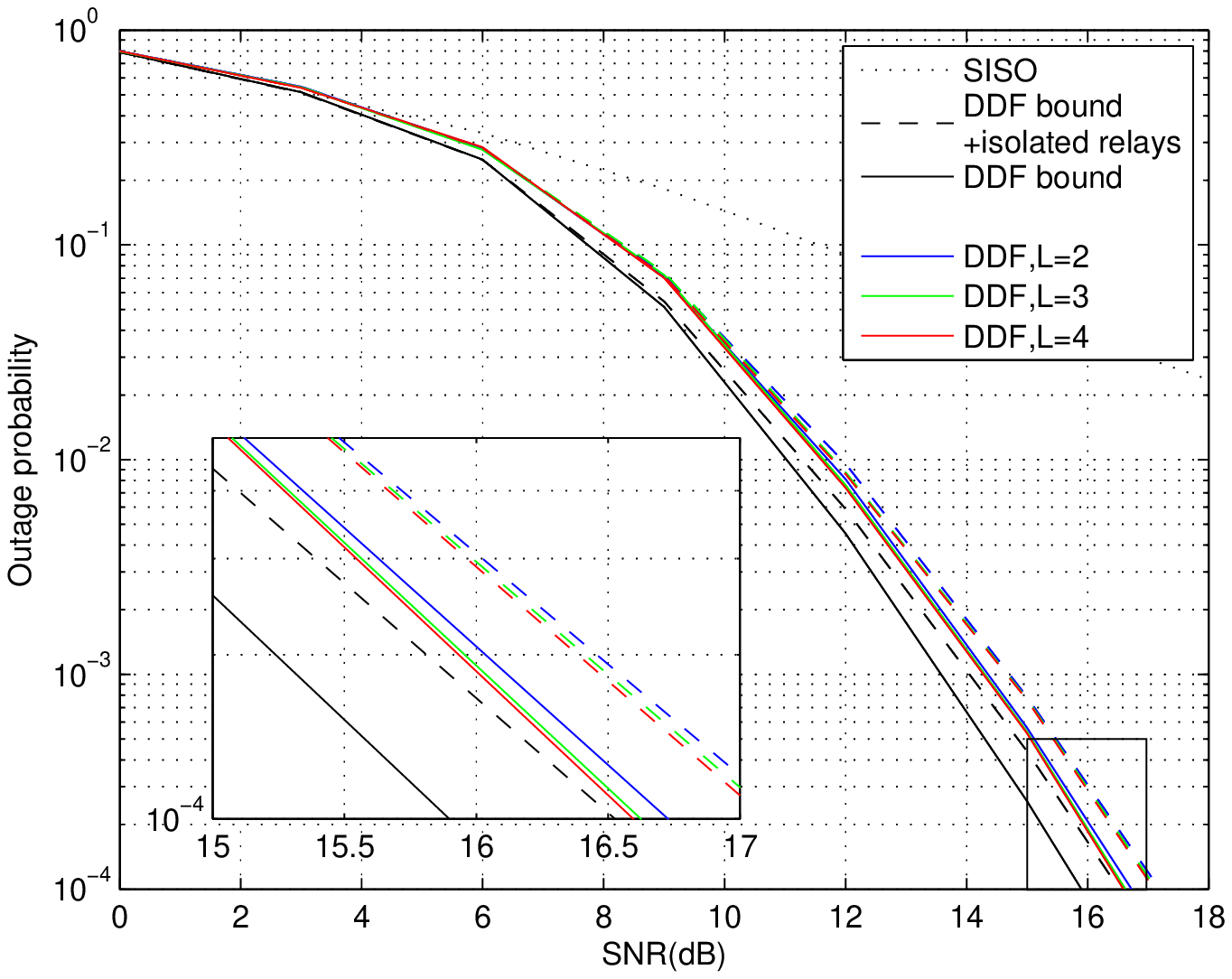}
 \caption{Outage probabilities of the DDF protocol implemented with different number of rotations for 3 relays with a frame-length of 64 symbols.}\label{pout_N3}
 \includegraphics[clip,width=\linewidth]{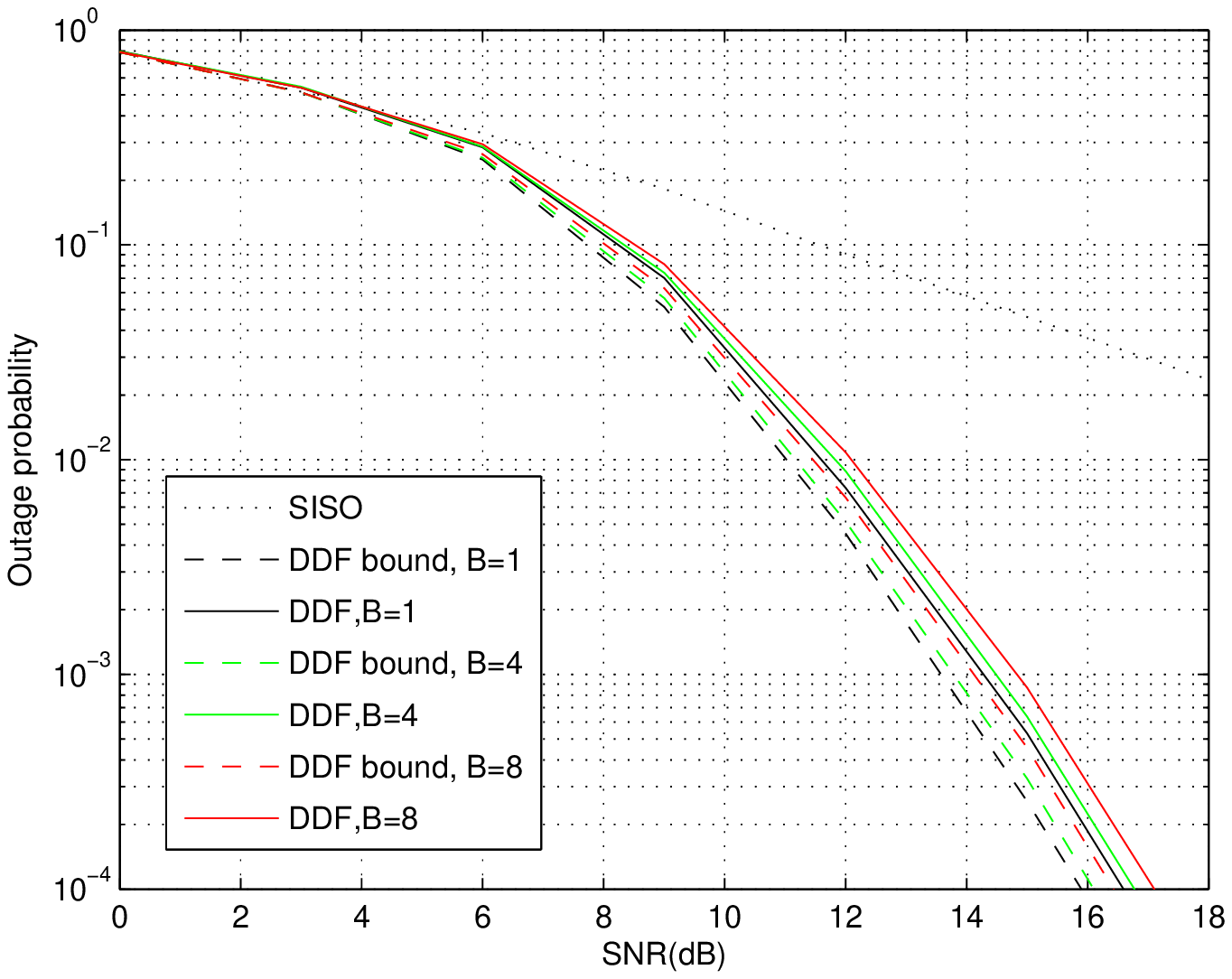}
 \caption{Outage probabilities of the DDF protocol implemented with 4 rotations for 3 relays and different block-lengths  with a frame-length of 64 symbols.}\label{pout_N3_block}
\end{figure}

In Fig. \ref{pout_N1}, the case of a single relay is studied. One can see that even a small number of rotations allows to reach performance close to the optimum one. (The optimal DDF performance is obtained by assuming an optimal $j\times1$ MISO transmission once $j$ relays have decoded).

In Fig. \ref{pout_N3}, the case of three relays is analyzed. The performance is plotted in two cases: either the relays are isolated ($\forall i,k, f_{ik} = 0$) or not. As expected, better performance are obtained when relays are communicating together. Indeed, more information is received at relays and thus they might be able to decode and retransmit earlier.

Previously, we assumed that relays attempt to decode at each time slot. The destination does not know when the relays will be able to decode, thus it has to assume the worst case where they have not all decoded till the last time slot. This means that at each time slot, relays have to broadcast at least $\lceil \log(N) \rceil$ additional bits indicating whether they have been able to decode or not. This way, the optimal performance is obtained, but the rate of the useful information is decreased. Thus, it has been suggested to divide the frame into blocks, and to allow relays to attempt decoding at the end of these blocks only \cite{azarian05}. The performance is then slightly decreased, but the rate is preserved. A tradeoff is introduced between performance and rate, which depends on the block-length $B$.

In Fig. \ref{pout_N3_block}, outage probabilities in the case of 3 relays are plotted with different block-lengths. One can see that the use of data blocks decreases the performance slightly, but the rates are considerably increased. Indeed, in this example, each symbol contains 2 bits. Since there are 3 relays, they need to send at least $\lceil \log(3) \rceil = 2$ bits before starting retransmission. Thus, in the case of no data block, the useful rate is only $R = \frac{2}{2+2} = \frac{1}{2}$ bpcu. In the case of blocks of length $B=4$, the useful rate is $R = \frac{2\times4}{2\times4+2} = \frac{4}{5}$ bpcu, and in the case of blocks of length $B=8$, it is $R = \frac{2\times8}{2\times8+2} = \frac{8}{9}$ bpcu. So the rate rapidly gets close to $1$ bpcu for a limited loss in performance.

\section{Conclusion}

In this paper, we have proposed a new implementation of the DDF adapted to any number of relays. This implementation is based on distributed rotations which allows to exploit space diversity without inducing extra complexity, since the decoding is similar to the one of a fast fading SISO channel.

The outage analysis, as well as the DMT computation in the single-relay two-rotation case, shows that this technique is very efficient even for a small number of rotations.

Moreover, the impact of block implementation is analyzed numerically, which shows that the loss in performance is limited compared to the considerable rate improvement.

In this paper, the order of rotation arrays is random. In a near future, authors plan to investigate the optimal order of rotations.

\section*{Acknowledgement}

This work was supported under Australian Research Council's Discovery Projects funding scheme (project no. DP0984950).

\bibliographystyle{IEEEtran}

\end{document}